\documentclass
[preprint,prb,superbib,superscriptaddress,byrevtex,showpacs]{revtex4}%
\usepackage{amsfonts}
\usepackage{amsmath}
\usepackage{amssymb}
\usepackage{graphicx}%
\begin{document}
\preprint{ }
\title[Short title for running header]{Study of one-dimensional nature of (Sr,Ba)$_{2}$Cu(PO$_{4}$)$_{2}$ and
BaCuP$_{2}$O$_{7}$ via $^{31}$P NMR. }
\author{R. Nath}
\affiliation{Department of Physics, Indian Institute of Technology, Mumbai 400076, India}
\author{A. V. Mahajan}
\affiliation{Department of Physics, Indian Institute of Technology, Mumbai 400076, India}
\author{N. B\"{u}ttgen}
\affiliation{Experimentalphysik V, Elektronische Korrelationen und Magnetismus, Institut
f\"{u}r Physik, Universit\"{a}t Augsburg, D-86135 Augsburg, Germany.}
\author{C. Kegler}
\affiliation{Experimentalphysik V, Elektronische Korrelationen und Magnetismus, Institut
f\"{u}r Physik, Universit\"{a}t Augsburg, D-86135 Augsburg, Germany.}
\author{A. Loidl}
\affiliation{Experimentalphysik V, Elektronische Korrelationen und Magnetismus, Institut
f\"{u}r Physik, Universit\"{a}t Augsburg, D-86135 Augsburg, Germany.}

\begin{abstract}
The magnetic behavior of the low-dimensional phosphates (Sr,Ba)$_{2}%
$Cu(PO$_{4}$)$_{2}$ and BaCuP$_{2}$O$_{7}$ was investigated by means of
magnetic susceptibility and $^{31}$P nuclear magnetic resonance (NMR)
measurements. We present here the NMR shift, the spin-lattice $\left(
1/T_{1}\right)  $ and spin-spin $\left(  1/T_{2}\right)  $ relaxation-rate
data over a wide temperature range $0.02$ K $\leq T\leq300$ K.\ The
temperature dependence of the NMR shift $K(T)$ is well described by the
$S=1/2$ Heisenberg antiferromagnetic chain model (D.C. Johnston \textit{et
al}., Phys. Rev. B \textbf{61}, 9558 (2000)) with an intrachain exchange of
$J/k_{B}$ $\simeq$ $165$ K, $151$ K, and $108$ K in Sr$_{2}$Cu(PO$_{4}$)$_{2}%
$, Ba$_{2}$Cu(PO$_{4}$)$_{2}$, and BaCuP$_{2}$O$_{7}$, respectively.
Deviations from Johnston's expression are seen for all these compounds in the
$T$-dependence of $K(T)$ at low temperatures. $^{31}$P is located
symmetrically between the Cu ions and fluctuations of the staggered
susceptibility at $q=\frac{\pi}{a}$ should be filtered out due to vanishing of
the geometrical form factor. However, the qualitative temperature dependence
of our $^{31}$P NMR $1/T_{1}$ indicates that relaxation due to fluctuations
around $q=\frac{\pi}{a}$ (but $\neq\frac{\pi}{a})$ have the same
$T$-dependence as those at $q=\frac{\pi}{a}$ and apparently dominate. Our
measurements suggest the presence of magnetic ordering at $0.8$ K in
BaCuP$_{2}$O$_{7}$ ($J/k_{B}\simeq108$ K) and a clear indication of a phase
transition (divergence) in $1/T_{1}(T)$, $1/T_{2}(T)$, and a change of the
line shape is observed. This enables us to investigate the 1D behavior over a
wide temperature range. We find that $1/T_{1}$ is nearly $T$-independent at
low-temperatures ($1$ K $\leq$ $T$ $\leq$ $10$ K), which is theoretically
expected for 1D chains when relaxation is dominated by fluctuations of the
staggered susceptibility. At high temperatures, $1/T_{1}$ varies nearly
linearly with temperature, which accounts for contribution of the uniform susceptibility.

\end{abstract}
\volumeyear{year}
\volumenumber{number}
\issuenumber{number}
\eid{identifier}
\date[Date text]{date}
\received[Received text]{date}

\revised[Revised text]{date}

\accepted[Accepted text]{date}

\published[Published text]{date}

\startpage{1}
\endpage{ }
\maketitle

\section{\textbf{INTRODUCTION}}

There is presently a lot of interest in the magnetic properties of
one-dimensional (1D) Heisenberg antiferromagnetic (HAF) spin systems. This is
due to the rich physics that they exhibit, in addition to the fact that such
systems are tractable from a computational/theoretical standpoint. In
particular, qualitative differences are expected between integer-spin and
half-integer spin HAF chains. While the integer-spin chains are
gapped,\cite{haldane1983} the half-integer spin chains are said to have quasi
long-range order (LRO) due to the gradual spatial decay (power-law) of the
spin-spin correlation function.\cite{bethe31, lieb61}

The magnetic Hamiltonian describing a spin-half Heisenberg chain can be
written as $H=-J\sum_{i}S_{i}\cdot S_{i+1}$ where $J$ is the intrachain
coupling constant between the nearest-neighbor spins. The temperature
dependence of the magnetic susceptibility $\chi\left(  T\right)  $ for the
$S=1/2$ HAF chain was numerically\ calculated by Bonner and Fisher
\cite{bonner1964} and since then, the Bonner-Fisher expression has been used
by experimentalists to determine the value of the exchange coupling $\left(
J\right)  $ from the temperature dependence of the bulk susceptibility. A more
accurate and analytical evaluation of the susceptibility of $S=1/2$ HAF chain
was done by Eggert, Affleck, and Takahashi \cite{eggert1994} which is valid at
low-temperatures. Via numerical simulations, an expression for $\chi\left(
T\right)  $ accurate for both low and high temperatures ($5\times10^{-25}%
\leq\frac{k_{B}T}{J}\leq5$, with $k_{B}$ the Boltzmann constant) was given by
Johnston {\textit{et al.}} \cite{johnston2000}. Dynamical properties of
$S=1/2$ chains have also been theoretically investigated. In particular, work
has been focussed on the properties measured by nuclear magnetic resonance
(NMR) techniques. Sachdev \cite{sachdev1994} determined the temperature
dependence of the NMR spin-lattice ($1/T_{1}$) and the Gaussian spin-spin
($1/T_{2G}$) relaxation rates for half-integer spin chains for $\frac{k_{B}%
T}{J}\ll1$ as: $1/T_{1}$ = constant and $1/T_{2G}$ $\propto$ $1/\sqrt{T}$.
Quantum Monte Carlo calculations by Sandvik\cite{sandvik1995} support these
results over an appropriate temperature range. These results are at variance
from those for classical spin chains ($S=\infty$), where theory
predicts\cite{hone1974} $1/T_{1}$ and $1/T_{2G}\propto T^{-3/2}$.

While the 1D compounds CuCl$_{2}$\textperiodcentered2NC$_{5}$H$_{5}$and
KCuF$_{3}$ have been experimentally investigated
previously,\cite{richards1974,chakhalian2003} the onset of LRO (at the
ordering temperature $T_{N}$) due to inter-chain interactions prevents the
study of true 1D properties down to low temperatures. For the abovementioned
1D compounds, the ratios $\frac{k_{B}T_{N}}{J}\simeq0.084$ and $0.1195$,
respectively, have been determined. Dynamic and static properties of Sr$_{2}%
$CuO$_{3}$ 1D chain have been extensively studied.
\cite{motoyama1996,takigawa1996,takigawa1997,thurber2001,barzykin2001} Due to
the large value of $J/k_{B}$ ($\approx$ $2200$ K) and the very weak
inter-chain couplings, this compound orders only below 5 K and hence 1D
properties could be studied in a large range of temperature. It is clearly
useful to examine new compounds which might exhibit 1D behavior in a large
temperature range, thereby allowing for a comparison with theoretical models
and improving our understanding of such systems.

Sr$_{2}$Cu(PO$_{4}$)$_{2}$ \cite{belik2002} and Ba$_{2}$Cu(PO$_{4}$)$_{2}$
\cite{etheredge1996} are two isostructural compounds having a monoclinic unit
cell with space group C$_{2/m}$. The reported lattice constants are $a=11.515$
\AA , $b=5.075$ \AA , $c=6.574$ \AA \ and $a=12.160$ \AA , $b=5.133$ \AA ,
$c=6.885$ \AA \ for Sr$_{2}$Cu(PO$_{4}$)$_{2}$ and Ba$_{2}$Cu(PO$_{4}$)$_{2}$,
respectively. BaCuP$_{2}$O$_{7}$, which differs slightly in structure compared
to the other two, crystallizes in a triclinic unit cell with space group
P$\overline{1}$ and lattice constants $a=7.353$ \AA , $b=7.578$ \AA ,
$c=5.231$ \AA .\cite{moqine1991} In the former two compounds, each CuO$_{4}$
\ square plane shares it's edges with two similar kind of PO$_{4}$ groups. The
edge sharing takes place in one direction forming an isolated [Cu(PO$_{4}%
$)$_{2}$]$_{\infty}$ chain along the crystallographic $b$-direction. A likely
interaction path of Cu$^{2+}$ ions is sketched in Fig. 1(a). As opposed to
Sr$_{2}$Cu(PO$_{4}$)$_{2}$ and Ba$_{2}$Cu(PO$_{4}$)$_{2}$, BaCuP$_{2}$O$_{7}$
contains two inequivalent P atoms, where each CuO$_{4}$ plaquette shares it's
edges with two different PO$_{4}$ groups forming chains as shown in Fig. 1(b).
Unlike the isolated chains of Sr$_{2}$Cu(PO$_{4}$)$_{2}$ and Ba$_{2}%
$Cu(PO$_{4}$)$_{2}$, there appear to be pairs of chains in BaCuP$_{2}$O$_{7}$.
Detailed magnetic properties of these compounds have not been reported yet.
Only Etheredge and Hwu\cite{etheredge1996} have published the bulk
susceptibility as a function of temperature for Ba$_{2}$Cu(PO$_{4}$)$_{2}$.
However, the authors failed to comment on the broad maximum at 80 K presumably
because it was suppressed by a high Curie contribution present in their sample.

In this paper we present in detail the magnetic properties of the 1D copper
phosphates, Sr$_{2}$Cu(PO$_{4}$)$_{2}$, Ba$_{2}$Cu(PO$_{4}$)$_{2}$, and
BaCuP$_{2}$O$_{7}$ using $^{31}$P NMR as a local probe. NMR is regarded as a
valuable tool for the study of microscopic properties of 1D chains, especially
through the studies of the NMR shift ($K$), the spin-spin relaxation rate
($1/T_{2}$), and the spin-lattice relaxation rate ($1/T_{1}$). We report on
measurements of the bulk susceptibility $\chi\left(  T\right)  $ for $1.8$ K
$\leq T\leq400$ K and $K(T)$, $1/T_{1}(T)$, and $1/T_{2}(T)$ of $^{31}$P NMR
in a large temperature range ($0.02$ K $\leq T\leq300$ K). The experimental
details concerning sample preparation and various measurements are given in
the next section. Section III contains our experimental results and a detailed
discussion about the results is presented in section IV. Our work on these
compounds strongly suggests that they are some of the best examples of $S=1/2$
1D HAF systems. In the course of our work, magnetic susceptibility and heat
capacity of Sr$_{2}$Cu(PO$_{4}$)$_{2}$, Ba$_{2}$Cu(PO$_{4}$)$_{2}$, and
BaCuP$_{2}$O$_{7}$ were reported by Belik \textit{et al.}%
\cite{belik2004,belik2004a}. They found the exchange constant ($J/k_{B}$) to
be $144$ K for Sr$_{2}$Cu(PO$_{4}$)$_{2}$, $132$ K for Ba$_{2}$Cu(PO$_{4}%
$)$_{2}$, and $103.6$ K for BaCuP$_{2}$O$_{7}$. Presence of any LRO was not
seen from specific heat measurement down to $0.45$ K for Sr$_{2}$Cu(PO$_{4}%
$)$_{2}$ and Ba$_{2}$Cu(PO$_{4}$)$_{2}$, whereas BaCuP$_{2}$O$_{7}$ showed
ordering at $0.81$ K.

\section{\textbf{EXPERIMENTAL DETAILS}}

Polycrystalline samples of Sr$_{2}$Cu(PO$_{4}$)$_{2}$, Ba$_{2}$Cu(PO$_{4}%
$)$_{2}$, and BaCuP$_{2}$O$_{7}$ were prepared by solid state reaction
techniques using BaCO$_{3}$ ($99.9$\% pure), SrCO$_{3}$ ($99.999$\% pure), CuO
($99.99$\% pure) and (NH$_{4})_{2}$HPO$_{4}$ ($99.9$\% pure) as starting
materials. The stoichiometric mixtures were fired at $800$ $^{\circ}$C
(Sr$_{2}$Cu(PO$_{4}$)$_{2}$), $700$ $^{\circ}$C (Ba$_{2}$Cu(PO$_{4}$)$_{2}$)
and $650$ $^{\circ}$C (BaCuP$_{2}$O$_{7}$) for $120$ hours each, in air, with
several intermediate grindings and pelletization. Finally some amounts of each
of the samples were annealed at $400$ $^{\circ}$C under a reducing atmosphere
($5$\% H$_{2}$ in Ar) in an attempt to reduce the Curie contribution in the
bulk susceptibility. Nearly single phases were confirmed from x-ray
diffraction, which was performed with a Philips\ Xpert-Pro powder
diffractometer. A Cu target was used in the diffractometer with $\lambda
_{av}=1.54182$ \AA . The x-ray diffraction patterns are shown in Fig. 2. Only
one compound (Sr$_{2}$Cu(PO$_{4}$)$_{2}$) exhibits minor impurity peaks which
are marked by asterisks. Lattice parameters were calculated using a
least-square fit procedure. The obtained lattice constants are ($a=11.496(5)$
\AA , $b=5.069(2)$ \AA , $c=6.566(3)$ \AA ), ($a=12.138(2)$ \AA , $b=5.123(1)$
\AA , $c=6.868(1)$ \AA ) and ($a=7.338(2)$ \AA , $b=7.561(2)$ \AA ,
$c=5.217(1)$ \AA ) for Sr$_{2}$Cu(PO$_{4}$)$_{2}$, Ba$_{2}$Cu(PO$_{4}$)$_{2}$,
and BaCuP$_{2}$O$_{7}$, respectively. These are in agreement with previously
reported values.

Magnetization ($M$) data were measured as a function of temperature $T$ ($1.8$
K $\leq$ $T$ $\leq$ $400$ K) and applied field $H$ ($0$ $\leq$ $H$ $\leq$ $70$
kG) using a SQUID magnetometer.

The NMR measurements were carried out using pulsed NMR techniques on $^{31}$P
nuclei (nuclear spin $I=1/2$ and gyromagnetic ratio $\gamma/2\pi$ = $17.237$
MHz/Tesla) in a large temperature range ($0.02$ K $\leq$ $T$ $\leq$ $300$ K).
We have done the measurements at two different applied fields of about $55$ kG
and $4$ kG which correspond to radio frequencies (rf) of about $95$ MHz and
$6.8$ MHz, respectively.

For $2$ K $\leq$ $T$ $\leq$ $300$ K, NMR measurements were done in a 55 kG
applied field with a $^{4}$He cryostat (Oxford Instruments). Spectra were
obtained by Fourier transform (FT) of the NMR echo signals using a $\pi/2$
pulse of width of about 4 $\mu s$. The NMR shift $K(T)=\left[  \nu\left(
T\right)  -\nu_{ref}\right]  /\nu_{ref}$ was determined by measuring the
resonance frequency of the sample ($\nu\left(  T\right)  $) with respect to a
standard H$_{3}$PO$_{4}$ solution (resonance frequency $\nu_{ref}$). The
spin-lattice relaxation rate $\left(  1/T_{1}\right)  $ was determined by the
inversion-recovery method. Spin-spin relaxation rate $\left(  1/T_{2}\right)
$ was obtained by measuring the decay of the echo integral with variable
spacing between the $\pi/2$ and the $\pi$ pulse.

In the $0.02$ K $\leq T\leq10$ K range, NMR measurements were performed using
a $^{3}$He/$^{4}$He dilution refrigerator (Oxford Instruments) with the
resonant circuit inside the mixing chamber. Spectra were obtained by field
sweeps at a constant radio frequency ($\nu_{rf}$) of $95$ MHz. $1/T_{1}$ was
measured down to $0.02$ K following the same procedure as described above
using $\pi/2$ pulse of width $15$ $\mu s$. Lower rf power (and consequently
longer pulse widths) were used to avoid rf heating of the sample. Measurements
were also done in a low-field of about $4$ kG ($\nu_{rf}$ $\simeq6.8$ MHz)
where the NMR line was narrow and inversion of the nuclear magnetization by a
$\pi$ pulse of width $30$ $\mu s$ was assured. The data from low-field
measurements almost reproduce the high-field data.

\section{\textbf{RESULTS}}

\subsection{Bulk Susceptibility}

Magnetic susceptibilities $\chi(T)$ (= $M/H$) for all the three compounds were
measured as a function of temperature in an applied field of 5 kG (Fig. 3).
The amount of ferromagnetic impurities present in our samples were estimated
from the intercept of $M$ vs. $H$ isotherms at various temperatures and were
found to be 19 ppm, 12 ppm, and 30 ppm of ferromagnetic Fe$^{3+}$ ions for
Sr$_{2}$Cu(PO$_{4}$)$_{2}$, Ba$_{2}$Cu(PO$_{4}$)$_{2}$, and BaCuP$_{2}$O$_{7}%
$, respectively. The data in Fig. 3 have been corrected for these
ferromagnetic impurities. As shown in the figure, all the samples exhibit a
broad maximum, indicative of low-dimensional magnetic interactions. With
decrease in temperature, susceptibility increases in a Curie-Weiss manner.
This possibly comes from chain ends, natural defects, excess oxygen and
extrinsic paramagnetic impurities present in the samples. No obvious features
associated with LRO are seen for $1.8$ K $\leq$ $T$ $\leq$ $400$ K for any of
the samples. A substantial reduction of Curie terms was achieved by annealing
the samples at 400 $^{0}$C in an atmosphere of 5\% H$_{2}$ in Ar. Similar
experiments in Sr$_{2}$CuO$_{3}$ and Y$_{2}$BaNiO$_{5}$ lead to reduced Curie
terms.\cite{ami1995, jaydip2003} In the insets of Figs. 3(a) and 3(b), data
for as-prepared and reduced Sr$_{2}$Cu(PO$_{4}$)$_{2}$ and Ba$_{2}$Cu(PO$_{4}%
$)$_{2}$, respectively are shown. Since the Curie contribution in the case of
as-prepared BaCuP$_{2}$O$_{7}$ is not large, we did not treat this sample in a
reducing atmosphere.

In order to fit the bulk susceptibility data, we assume that the
susceptibility consists of three terms:

\textit{%
\begin{equation}
\chi=\chi_{0}+\frac{C}{T+\theta}+\chi_{spin}(T)
\end{equation}
}

where $\chi_{spin}(T)$ is the uniform spin susceptibility for a $S=1/2$ 1D HAF
system given in Ref. 6 (expression corresponding to \textquotedblleft
fit2\textquotedblright). This expression (containing the the Land\'{e}
$g$-factor and $J$ as fitting parameters) is not reproduced here since it is
somewhat unwieldy.The first term $\chi_{o}$ is temperature independent and
consists of diamagnetism of the core electron shells ($\chi_{core}$) and
Van-Vleck paramagnetism ($\chi_{vv}$) of the open shells of the Cu$^{2+}$ ions
present in the sample. The second term $\frac{C}{T+\theta}$ is the Curie-Weiss
contribution due to paramagnetic species in the sample.

The average Land\'{e} $g$-factors determined from an analysis of the powder
spectra from electron paramagnetic resonance (EPR) experiments on our samples
were found to be 2.15, 2.15, and 2.2 for Sr$_{2}$Cu(PO$_{4}$)$_{2}$, Ba$_{2}%
$Cu(PO$_{4}$)$_{2}$, and BaCuP$_{2}$O$_{7}$, respectively. Our experimental
$\chi(T)$\ data were fitted using the above $g$-values (the solid lines are
the best fits in Fig. 3) and the extracted parameters are listed in Table I.

Table I.%

\begin{tabular}
[c]{ccccc}\hline\hline
Sample & $\chi_{0}$ & $C$ & $\theta$ & $\frac{J}{k_{B}}$\\\hline
& $10^{-3}$ cm$^{3}$/mole & $10^{-3}$ cm$^{3}$K/mole & K & K\\\hline
Sr$_{2}$Cu(PO$_{4}$)$_{2}$ & $0.005$ & $10.6$ & $1.1$ & $152$\\
Ba$_{2}$Cu(PO$_{4}$)$_{2}$ & $-0.15$ & $6.8$ & $0.5$ & $146$\\
BaCuP$_{2}$O$_{7}$ & $-0.07$ & $1.6$ & $0.4$ & $108$\\\hline\hline
\end{tabular}

\bigskip

Adding the core diamagnetic susceptibility for the individual
ions\cite{Magnetochemistry}, the total $\chi_{core}$ was calculated to be
$-1.39\times10^{-4}$ cm$^{3}$/mole, $-1.73\times10^{-4}$ cm$^{3}$/mole, and
$-1.29\times10^{-4}$ cm$^{3}$/mole for Sr$_{2}$Cu(PO$_{4}$)$_{2}$, Ba$_{2}%
$Cu(PO$_{4}$)$_{2}$, and BaCuP$_{2}$O$_{7}$, respectively. The Van-Vleck
paramagnetic susceptibility for our samples estimated by subtracting
$\chi_{core}$ from $\chi_{0}$ gives $\chi_{vv}=14.4\times10^{-5}$ cm$^{3}%
$/mole, $2.3\times10^{-5}$ cm$^{3}$/mole, and $5.9\times10^{-5}$ cm$^{3}$/mole
for Sr$_{2}$Cu(PO$_{4}$)$_{2}$, Ba$_{2}$Cu(PO$_{4}$)$_{2}$, and BaCuP$_{2}%
$O$_{7}$, respectively. These values are comparable to that found in Sr$_{2}%
$CuO$_{3}$\cite{motoyama1996}. The Curie contributions present in the samples
correspond to a defect spin concentration of $3$ \%, $1.8$ \% and $0.4$ \% for
Sr$_{2}$Cu(PO$_{4}$)$_{2}$, Ba$_{2}$Cu(PO$_{4}$)$_{2}$, and BaCuP$_{2}$%
O$_{7},$ respectively assuming defect spin $S=1/2$.

\subsection{$^{31}$P NMR}

\subsubsection{NMR shift}

NMR has an important advantage over bulk susceptibility for the determination
of magnetic parameters. While the presence of a Curie-like contribution
restricts the accurate determination of $\chi_{Spin}(T)$ and hence $J$ from
$\chi(T)$, in NMR this paramagnetism broadens the NMR line but does not
contribute to the NMR\ shift $K$. Therefore, it is more reliable to extract
the $\chi_{Spin}(T)$ and $J$ from the temperature dependence of the NMR shift
rather than from the bulk susceptibility. From Fig.1 it appears that in all
the compounds each $^{31}$P is coupled to two Cu$^{2+}$ ions via a
super-transferred hyperfine interaction mediated by oxygen ions in its
neighborhood. All the NMR data reported in this paper correspond to samples of
Sr$_{2}$Cu(PO$_{4}$)$_{2}$ and Ba$_{2}$Cu(PO$_{4}$)$_{2}$ which were treated
in a reducing atmosphere as described in the previous section while the data
for BaCuP$_{2}$O$_{7}$ correspond to the as-prepared sample. We note here that
we also did NMR measurements on the as-prepared Sr$_{2}$Cu(PO$_{4}$)$_{2}$ and
Ba$_{2}$Cu(PO$_{4}$)$_{2}$ samples (for $T$
%TCIMACRO{\TEXTsymbol{>} }%
%BeginExpansion
$>$
%EndExpansion
$10$ K) and found no differences with respect to the samples which were
treated in a reducing atmosphere.

NMR shift data as a function of temperature are shown in Fig. 4. The samples
exhibit broad maxima at different temperatures: $\approx$ $100$ K for Sr$_{2}%
$Cu(PO$_{4}$)$_{2}$, $\approx$ $90$ K for Ba$_{2}$Cu(PO$_{4}$)$_{2}$, and
$\approx$ $70$ K for BaCuP$_{2}$O$_{7}$, indicative of short-range ordering.
Towards lower temperatures $T$
%TCIMACRO{\TEXTsymbol{<} }%
%BeginExpansion
$<$
%EndExpansion
$20$ K, the NMR shift $K(T)$ shows a plateau as is demonstrated by the
semilogarithmic plot in the inset of Fig. 4. In the sub-Kelvin region, the NMR
shift $K(T)$ of all our samples decreases steeply. The fall-off appears below
$k_{B}T/J\simeq0.003$ for Sr$_{2}$Cu(PO$_{4}$)$_{2}$, $k_{B}T/J\simeq0.0033$
for Ba$_{2}$Cu(PO$_{4}$)$_{2}$, and $k_{B}T/J\simeq0.01$ for BaCuP$_{2}$%
O$_{7}$.

The conventional scheme of analysis is to first determine the spin
susceptibility (as done in the previous section) and then plot $K$ vs
$\chi_{Spin}$ with $T$ as an implicit parameter. The exchange coupling $J$ is
obtained from the susceptibility analysis while the slope of the $K$ vs
$\chi_{Spin}$ plot yields the total hyperfine coupling $A$ between the $^{31}%
$P nucleus and the two nearest-neighbor Cu$^{2+}$ ions. As an example, such a
$K-\chi_{spin}$ plot is shown in Fig. 5 for BaCuP$_{2}$O$_{7}$.

Since an algebraic expression for the temperature dependence of the spin
susceptibility (and therefore the spin-shift) is known in this case, we prefer
to determine $J$ and $A$ simultaneously by fitting the temperature dependence
of $K$ to the following equation,

\textit{%
\begin{equation}
K=K_{0}+\left(  \frac{A}{N_{A}\mu_{B}}\right)  \chi_{spin}\left(  T,J\right)
\end{equation}
}

where $K_{0}$ is the chemical shift and $N_{A}$ is the Avogadro number. While
fitting, $g$ was kept fixed to the value obtained from EPR analysis and
$K_{0}$, $A$, and $J$ were free parameters. The parameters $J$ and $A$
determined in this manner are considered more reliable, since the only
temperature dependent term in the NMR shift is due to spin-susceptibility,
while bulk susceptibility analysis is somewhat hampered by low-temperature
Curie terms. As shown in Fig. 4, the shift data fit nicely to Eq. 2 in the
temperature range $10$ K $\leq$ $T$ $\leq$ $300$ K yielding the parameters
given in Table II.

Table II.%

\begin{tabular}
[c]{llll}\hline\hline
Sample & $K_{0}$ & $A$ & $J/k_{B}$\\\hline
& ppm & Oe/$\mu_{B}$ & K\\\hline
Sr$_{2}$Cu(PO$_{4}$)$_{2}$ & $47$ & $2609\pm100$ & $165$\\
Ba$_{2}$Cu(PO$_{4}$)$_{2}$ & $40$ & $3364\pm130$ & $151$\\
BaCuP$_{2}$O$_{7}$ & $73$ & $2182\pm20$ & $108$\\\hline\hline
\end{tabular}

\subsubsection{Spectra}

For all the three compounds the $^{31}$P NMR spectra consist of a single
spectral line as is expected for $I$ = $1/2$ nuclei (Fig. 6). As shown in the
crystal structures in Fig. 1, Sr$_{2}$Cu(PO$_{4}$)$_{2}$ and Ba$_{2}%
$Cu(PO$_{4}$)$_{2}$ have a unique $^{31}$P site, whereas in BaCuP$_{2}$O$_{7}$
there are two inequivalent $^{31}$P sites. \ However, a single resonance line
even for BaCuP$_{2}$O$_{7}$ implies that both the $^{31}$P sites in this
compound are nearly identical. Since our measurements are on randomly oriented
polycrystalline samples, asymmetric shape of the spectra corresponds to a
powder pattern due to an asymmetric hyperfine coupling constant and an
anisotropic susceptibility. The linewidth was found to be field and
temperature dependent as is shown in the insets of Figs. 6(a) and (b) for
(Sr/Ba)$_{2}$Cu(PO$_{4}$)$_{2}$ and in Figs. 6(c) and (d) for BaCuP$_{2}%
$O$_{7}$, respectively. While we did not do a detailed analysis of the
linewidth, its $T$- and $H$-dependence is likely due to macroscopic field
inhomogeneities due to the demagnetization effects of a powder
sample\cite{lomer1962} and paramagnetic impurities.

As seen from Fig. 6 (c) the NMR spectra of BaCuP$_{2}$O$_{7}$ broaden abruptly
below about $0.8$ K. We then measured the spectral lineshape in a low field
($H$ $\simeq4$ kG) below $0.8$ K in order to check whether any features could
be resolved. Fig. 6 (d) shows the appearance of two shoulders on either side
of the central line below $0.8$ K. This is most likely an indication of the
appearance of LRO. The positions of the shoulders stay unchanged with
temperature while their relative intensity increases with decreasing
temperature. A more detailed discussion is carried out in section IV.

\subsubsection{Spin-lattice relaxation rate $1/T_{1}$}

Temperature dependencies of $^{31}$P $1/T_{1}$ are presented in Fig.7. For the
$1/T_{1}$ experiment, the central positions of corresponding spectra at high
($55$ kG) and low ($4$ kG) external fields have been irradiated. Inset of
Fig.7 (a) shows the typical magnetization recovery at $H$ $\simeq$ $55$ kG and
at two different temperatures. For a spin-1/2 nucleus the recovery is expected
to follow a single exponential behavior. In Sr$_{2}$Cu(PO$_{4}$)$_{2}$ and
Ba$_{2}$Cu(PO$_{4}$)$_{2}$ (for $H$ $\simeq$ $55$ kG), the recovery of the
nuclear magnetization after an inverting pulse was single exponential down to
$2$ K while for $T$ $<$ $2$ K, it fitted well to the double exponential,%

\begin{equation}
\frac{1}{2}\left(  \frac{M\left(  \infty\right)  -M\left(  t\right)
}{M\left(  \infty\right)  }\right)  =A_{1}\exp\left(  -\frac{t}{T_{1L}%
}\right)  +A_{2}\exp\left(  -\frac{t}{T_{1S}}\right)  +C
\end{equation}
where $1/T_{1L}$ corresponds to the slower rate and $1/T_{1S}$ is the faster
component. $M(t)$ is the nuclear magnetization a time $t$ after an inverting
pulse. \ Since the deviation from single exponential behavior could be due to
the large linewidth and our consequent inability to saturate the NMR line, we
also performed experiments at a lower field ($\simeq$ $4$ kG), where the line
is about three times narrower (see insets of Figs. 6 (a) or (b)). However,
even in low-field (where the rf field $H_{1}$ was sufficient to ensure
complete inversion) the nuclear magnetization recovery is not single
exponential implying that this is an intrinsic effect. With increasing
temperature, the ratio $\frac{A_{2}}{A_{1}}$ decreases and the recovery
becomes single exponential for $T$ $>$ $2$ K. It appears that the longer
$T_{1}$ component comes from the chain itself while the faster component is
associated with $^{31}$P nuclei near chain ends. Clearly, at lower
temperatures, the chain-end-induced magnetization extends to large distances
from chain ends (thereby affecting more $^{31}$P nuclei) and consequently the
weight associated with the faster relaxation is more at lower temperatures.
From the experiment, it was found that low-field measurements reproduce almost
the same $T_{1}$ as for high-field. \cite{footnote1} Although we have measured
$T_{1}$ down to $0.02$ K, since the magnetization recovery is not
single-exponential, reliable relaxation rates $1/T_{1}$ couldn't be obtained
below $0.1$ K (where the faster component $1/T_{1s}$ has a large weight). Fig.
7 (a) and (b) display data down to $0.1$ K, where it is seen that $1/T_{1}$
for Sr$_{2}$Cu(PO$_{4}$)$_{2}$ and Ba$_{2}$Cu(PO$_{4}$)$_{2}$ do not show any
anomaly. \ Even at lower temperatures, there was no indication of a divergence
of the relaxation rate, indicating the absence of any magnetic ordering. For
$1$ K $\leq$ $T$ $\leq$ $10$ K, $1/T_{1}$ remains constant with temperature
and below $0.5$ K a slight increase was observed for both Sr$_{2}$Cu(PO$_{4}%
$)$_{2}$ and Ba$_{2}$Cu(PO$_{4}$)$_{2}$. At high temperatures ($T$ $\geq$ $30$
K), $1/T_{1}$ varies nearly linearly with temperature.

In BaCuP$_{2}$O$_{7}$, $1/T_{1}$ at $H$ $\simeq$ $55$ kG was measured down to
3 K. Once again, the large line width prevented us from saturating the nuclear
magnetization below 3 K. Low-field measurements give perfect single
exponential recovery down to 2 K and below 2 K, it was fitted well to double
exponential. From Fig. 7(c), it is clear that the $1/T_{1}(T)$ diverges at $T$
$\approx$ $0.8$ K, indicating an approach to magnetic ordering. For $1$ K
$\leq$ $T$ $\leq$ $10$ K, $1/T_{1}$ remains constant and for $T$ $\geq15$ K,
it varies linearly with temperature. Slight change in magnitude in low-field
data compared to high-field data may be due to
spin-diffusion.$\cite{takigawa1996}$

\subsubsection{Spin-spin relaxation rate $1/T_{2G}$}

Spin-spin relaxation was measured at $H$ $\simeq$ $4$ kG, where the line is
sufficiently narrow. The spin-spin relaxation was found to have a Gaussian
behavior and the rate ($1/T_{2G}$) was obtained by monitoring the decay of the
transverse magnetization after a $\frac{\pi}{2}-t-\pi$ pulse sequence, as a
function of the pulse separation time $t$, and fitting to the following equation,%

\begin{equation}
M\left(  2t\right)  =M_{0}\exp[-2\left(  \frac{t}{T_{2G}}\right)  ^{2}]+C
\end{equation}

As shown in the inset of Fig.8, the spin-spin relaxation rate $1/T_{2G}$ for
all the samples is nearly temperature independent. BaCuP$_{2}$O$_{7}$ is the
only compound which exhibits a significantly enhanced spin-spin relaxation
rate $1/T_{2G}$ at the lowest temperature compared to elevated temperatures.
This increase of $1/T_{2G}$ in BaCuP$_{2}$O$_{7}$ is most likely related to LRO.

\section{DISCUSSION}

\subsection{NMR shift}

The general variation of the shift with temperature follows the expected
behavior of an $S=1/2$ HAF chain, as seen in the results section. A steep
decrease in $K(T)$ was observed below $k_{B}T/J\simeq0.003$ for Sr$_{2}%
$Cu(PO$_{4}$)$_{2}$, $k_{B}T/J\simeq0.0033$ for Ba$_{2}$Cu(PO$_{4}$)$_{2}$ and
$k_{B}T/J\simeq0.01$ for BaCuP$_{2}$O$_{7}$.\cite{footnote2} This decrease of
$K$ is clearly much more than the logarithmic decrease with an infinite slope
at zero temperature, expected from theory (see solid line in Fig. 4). In
Sr$_{2}$Cu(PO$_{4}$)$_{2}$, Ba$_{2}$Cu(PO$_{4}$)$_{2}$, and BaCuP$_{2}$O$_{7}%
$, from Fig. 4, the extrapolated shifts at zero temperature were found to be
400 ppm, 590 ppm, and 580 ppm, respectively. The theoretically expected
values, derived from the $T$ $=0$ susceptibility\cite{footnote3} and using the
relevant $A$ and $K_{0}$, are 544 ppm, 740 ppm, and 738 ppm, respectively for
Sr$_{2}$Cu(PO$_{4}$)$_{2}$, Ba$_{2}$Cu(PO$_{4}$)$_{2}$, and BaCuP$_{2}$O$_{7}%
$. Among the various causes for this deviation, one might be the onset of
spin-Peierls order. In such a case, the spin susceptibility (and therefore the
spin shift) should vanish at $T=0$. However, our extrapolated $T=0$ shifts are
much more than the chemical shifts $K_{0}$ and there is no exponential
decrease of $1/T_{1}(T)$ toward low temperatures. Another possibility is the
onset of 3D LRO. In this case, a divergence should have been seen in the
temperature dependencies of the spin-lattice relaxation rate $1/T_{1}$ as well
as in the spin-spin relaxation rate $1/T_{2G}$. While this is the case for
BaCuP$_{2}$O$_{7}$, only a small increase of the relaxation rates is observed
for Sr$_{2}$Cu(PO$_{4}$)$_{2}$ and Ba$_{2}$Cu(PO$_{4}$)$_{2}$. A clear effect
is observed in the temperature dependencies of $K(T)$, $1/T_{1}$, $1/T_{2}$,
and lineshape for BaCuP$_{2}$O$_{7}$ at $0.8$ K. This establishes the onset of
LRO at $0.8$ K in BaCuP$_{2}$O$_{7}$. However, in the case of Sr$_{2}%
$Cu(PO$_{4}$)$_{2}$ and Ba$_{2}$Cu(PO$_{4}$)$_{2}$, while a clear anomaly is
seen in $K(T)$ at low-temperature, only a weak anomaly is seen in $1/T_{1}(T)$
and no significant changes were observed either in the low-temperature spectra
or in $1/T_{2}(T)$. In summary, the presence or absence of LRO at
low-temperatures in Sr$_{2}$Cu(PO$_{4}$)$_{2}$ and Ba$_{2}$Cu(PO$_{4}$)$_{2}$,
can't be unambiguously established.

\subsection{Effect of chain ends on the NMR spectrum}

From field-theory and Monte Carlo calculations, Eggert and
Affleck\cite{eggert1995} found that in case of half-integer spin chains, the
local susceptibility near an open end of a finite chain has a large
alternating component. This component appears in the form of staggered
magnetization near chain ends, under the influence of an uniform field. This
staggered moment has a maximum at a finite distance from the end and increases
as $1/T$ with decreasing temperature. Analytical expression for the spin
susceptibility $\chi\left(  l\right)  $ at site $l$ were obtained which
consist of the uniform ($\chi_{u}$) and alternating ($\chi_{alt}$)
parts$\cite{eggert1995}$ $\chi\left(  l\right)  =\chi_{u}\left(  l\right)
+\left(  -1\right)  ^{l}\chi_{alt}\left(  l\right)  ,$ where the uniform part
is nearly constant and the alternating part is given by%

\begin{equation}
\chi_{alt}\left(  l\right)  =\frac{aJ}{v}\frac{l}{\sqrt{\left(  v/\pi
T\right)  \sinh\left(  2\pi Tl/v\right)  }}%
\end{equation}
\bigskip

where $v=\pi J/2$ is the spin-wave velocity. The NMR spectrum represents the
distribution function of NMR shift, which is equivalent to $\chi_{alt}$, and
has the form $g\left(  x\right)  =\sum_{l}f\left[  x-\left(  -1\right)
^{l}\chi_{alt}\left(  l\right)  \right]  $, where $f$ takes the form of a
Lorentzian. This expression is valid for in-chain Cu$^{2+}$ site. In our
compounds, $^{31}$P is the probe nucleus, which is sensitive to two nearest
neighbor Cu$^{2+}$ ions belonging to one chain. $^{31}$P NMR line shape is
then given by $g\left(  x\right)  =\sum_{l}f\left[  x-\{\left(  -1\right)
^{l}\chi_{alt}\left(  l\right)  +\left(  -1\right)  ^{l+1}\chi_{alt}\left(
l+1\right)  \}\right]  $. Fig. 9 shows the simulated spectra for both Cu and P
sites choosing $\ f$ to be a Lorentzian with width 0.05.

For $k_{B}T=J/33$, at Cu site, $\chi_{alt}$ has a maximum at $l=0.48J/T$,
which results in features in the spectra on either side of the central line.
With increasing temperature these features on either side of the central peak
disappear and $\sqrt{T}(\frac{\Delta H}{2H_{0}})$ (where $\Delta H$ is the
width at the background and H$_{0}$ is the field at the central peak) remain
constant with temperature. This has been seen by Takigawa \cite{takigawa1997a}
in $^{63}$Cu NMR spectra of Sr$_{2}$CuO$_{3}$. However, the static effects of
the staggered magnetization at the $^{31}$P site in our compounds are expected
to be\ much weaker (Fig. 9) due to the near cancellation of the magnetization
from the neighboring Cu sites.

Our low-temperature spectra for Sr$_{2}$Cu(PO$_{4}$)$_{2}$ and Ba$_{2}%
$Cu(PO$_{4}$)$_{2}$ show a single spectral line without any shoulders on
either sides of the central peak. This is in agreement with our expectations
that we are not sensitive to chain-end effects in $^{31}$P NMR spectra\ of
these compounds.

In low-field NMR on BaCuP$_{2}$O$_{7}$, a sudden increase of linewidth was
observed below $0.8$ K along with appearance of two shoulder-like features
located symmetrically on either side of the central peak (Fig.6(d)). If these
shoulder-like features come from the staggered magnetization of chain ends
then, as discussed above, $\sqrt{T}(\frac{\Delta H}{2H_{0}})$ should be
temperature independent with the shoulders moving outwards at lower
temperatures. But in BaCuP$_{2}$O$_{7}$ the shoulder positions are temperature
independent, suggesting that those features are not chain-end effects.
Further, in case of a structural phase transition, symmetrically located
features which become more intense as temperature is lowered are not expected.
Our data are consistent with an alternating magnetization in the Cu chain.
Further, the magnitude of this magnetization must have a maximum away from the
chain end (so as to produce the symmetric features) and the location (site
$l_{max}$) of the maximum must not change with $T$ (as the shoulder positions
are temperature independent). As shown before, the chain-end induced effects
are expected to be negligible at $^{31}$P site. The large effect seen here
must stem from a magnetic transition possibly with exotic spin-order.

\subsection{Wave vector $\mathbf{q}$\textbf{-} and temperature $T$- dependence
of $1/T_{1}$}

In order to study the microscopic behavior of 1D HAF systems, it is useful to
measure the temperature dependence of spin-lattice relaxation rate which
yields information on the imaginary part of the dynamic susceptibility
$\chi\left(  \mathbf{q,\omega}\right)  $. The spin-lattice relaxation rate, in
general, is affected by both uniform $\left(  q=0\right)  $ and staggered spin
fluctuations $\left(  q=\pm\frac{\pi}{a}\right)  $. The uniform component
leads to $1/T_{1}\propto T$, while the staggered component gives $1/T_{1}$ =
constant.\cite{sachdev1994} At the $^{31}$P sites, $q$-dependence of $1/T_{1}$
can be expressed in terms of form factors as,%

\begin{equation}
1/T_{1}\propto\sum_{q}\left[  A^{2}\cos^{2}\left(  qx\right)  \right]
\operatorname{Im}\chi\left(  q,\omega\right)
\end{equation}

We are probing on the $^{31}$P nucleus, which is linked to the Cu spins via
O(1) and O(2) as shown in Fig. 1. Since $^{31}$P is symmetrically located
between the Cu ions, the antiferromagnetic fluctuations are filtered at the
$^{31}$P site, provided the two hyperfine couplings are equal. However,
contributions just slightly differing from $q=\pi/a$ are apparently strong
enough and/or the two hyperfine couplings are unequal and result in a
qualitative behavior expected when relaxation is dominated by fluctuations of
the staggered susceptibility.

If one were to ignore the geometrical form factor completely, the relaxation
rate due to staggered fluctuations can be calculated following the
prescription of Barzykin\cite{barzykin2001}. He obtained the normalized
dimensionless NMR spin-lattice relaxation rate at low-temperature $\left(
1/T_{1}\right)  _{norm}=\frac{\hbar J}{A_{th}^{2}T_{1}}\approx0.3$, where
$A_{th}$ is $A\left(  2h\gamma/2\pi\right)  $. Assuming the fluctuations to be
correlated, $1/T_{1}$ can be written as $1/T_{1}=\frac{0.3}{\hbar J/A^{2}}$.
Using this expression, $\left(  1/T_{1}\right)  $ at the $^{31}$P site
was\ calculated to be about $44$ sec,$^{-1}$ $80$ sec,$^{-1}$ and $47$
sec,$^{-1}$ whereas our experimental values are $15$ sec,$^{-1}$ $20$
sec,$^{-1}$ and $25$ sec,$^{-1}$ for Sr$_{2}$Cu(PO$_{4}$)$_{2}$, Ba$_{2}%
$Cu(PO$_{4}$)$_{2}$, and BaCuP$_{2}$O$_{7}$ respectively in the $1$ K $\leq T$
$\leq$ $10$ K range. The experimental values are clearly smaller than the
theoretical ones due to the geometrical form factor. Further, a logarithmic
increase of $1/T_{1}$ is expected at low temperatures
following\cite{barzykin2001}
\begin{equation}
\left(  1/T_{1}\right)  _{norm}=2D\sqrt{\ln\frac{\Lambda}{T}+\frac{1}{2}%
\ln\left(  \ln\frac{\Lambda}{T}\right)  }\left(  1+O\left[  \frac{1}{\ln
^{2}\frac{\Lambda}{T}}\right]  \right)
\end{equation}
where $D=1/(2\pi)^{3/2}$, $\Lambda$ is the cutoff parameter given by
$2\sqrt{2\pi}e^{C+1}J$ and $C$ ($\simeq0.5772157)$ is Euler's constant. In
Fig. 10, we have plotted $\left(  (1/T_{1})/(1/T_{1})_{T=10K}\right)  _{norm}$
calculated from equation 7 and the experimental results of the spin-lattice
relaxation rate $1/T_{1}$ for our compounds normalised by their values at $10$
K, as a function of temperature. The qualitative temperature dependence of the
experimental spin-lattice relaxation rate $1/T_{1}$ agrees reasonably well
with theoretical calculations in the constant region but below $0.5$ K, the
experimental increase in Sr$_{2}$Cu(PO$_{4}$)$_{2}$ and Ba$_{2}$Cu(PO$_{4}%
$)$_{2}$ is somewhat more than the logarithmic increase expected
theoretically. However, complementary measurements are needed to fully
understand this issue.

\subsection{Spin-spin relaxation rate $1/T_{2G}$}

Following the treatment of Sachdev\cite{sachdev1994} and
Barzykin,\cite{barzykin2001} spin-spin relaxation is expected to follow a
Gaussian behavior with a temperature dependence given by $1/T_{2G}%
\propto1/\sqrt{T}$. Further, the spinon mediated spin-spin relaxation rate can
be calculated\cite{barzykin2001} as follows. The normalized spin-spin
relaxation rate is given by $\left(  \frac{\sqrt{T}}{T_{2G}}\right)
_{norm}=\left(  \frac{k_{B}T}{J}\right)  ^{1/2}\frac{\hbar J}{A^{2}T_{2G}}$.
Dividing $\left(  \frac{1}{T_{1}}\right)  _{norm}$ by $\left(  \frac{\sqrt{T}%
}{T_{2G}}\right)  _{norm}$ and equating it to 1.8 (the value obtained from
Ref. 16) one finds that $1/T_{2G}=\frac{44}{1.8}\left(  \frac{J}{k_{B}%
T}\right)  ^{1/2}$, $\frac{80}{1.8}\left(  \frac{J}{k_{B}T}\right)  ^{1/2}$,
and $\frac{47}{1.8}\left(  \frac{J}{k_{B}T}\right)  ^{1/2}$, respectively for
Sr$_{2}$Cu(PO$_{4}$)$_{2}$, Ba$_{2}$Cu(PO$_{4}$)$_{2}$, and BaCuP$_{2}$O$_{7}%
$. This leads to T$_{2G}$ values $3.2$ ms, $1.8$ ms, and $3.6$ ms in contrast
to our experimental values $269$ $\mu s$, $255$ $\mu s$, and $207$ $\mu
s$\ respectively for Sr$_{2}$Cu(PO$_{4}$)$_{2}$, Ba$_{2}$Cu(PO$_{4}$)$_{2}$,
and BaCuP$_{2}$O$_{7}$ at $1$ K. These are almost three orders of magnitude
smaller than the theoretically calculated values. Further, our experimental
spin-spin relaxation rates are temperature independent. Clearly, in the
present case, spinon mediated coupling does not contribute to spin-spin
relaxation. On the other hand, an estimate of the nuclear dipole-dipole
mediated relaxation $\left(  T_{2}=\frac{r^{3}}{\gamma^{2}\hbar}\text{, where
}r\text{ is the dipole-dipole distance}\right)  $ would seem to explain the
observed relaxation rates. This must be primarily because of the small
exchange coupling in contrast to Sr$_{2}$CuO$_{3}$ where $J$ is an order of
magnitude larger and spinon mediated Gaussian spin-spin relaxation rate has
been observed.

\section{\textbf{CONCLUSION}}

Our NMR and susceptibility measurements on Sr$_{2}$Cu(PO$_{4}$)$_{2}$,
Ba$_{2}$Cu(PO$_{4}$)$_{2}$, and BaCuP$_{2}$O$_{7}$ show good agreement with
the theory of 1D $S=1/2$ Heisenberg antiferromagnetic chains. NMR shift $K$as
a function of temperature fitted well to the recent theoretical calculation by
Johnston and the exchange interaction $J/k_{B}$ is estimated to be $165$ K,
$151$ K, and $108$ K for Sr$_{2}$Cu(PO$_{4}$)$_{2}$, Ba$_{2}$Cu(PO$_{4}$%
)$_{2}$, and BaCuP$_{2}$O$_{7},$ respectively. We observed a steep decrease of
the NMR shift $K(T)$ below $T\simeq0.003J/k_{B}$, $0.0033J/k_{B}$, and
$0.01J/k_{B}$ for Sr$_{2}$Cu(PO$_{4}$)$_{2}$, Ba$_{2}$Cu(PO$_{4}$)$_{2}$, and
BaCuP$_{2}$O$_{7}$, respectively. Low-field $^{31}$P NMR spectra of
BaCuP$_{2}$O$_{7}$ shows sudden appearance of broad humps on either side of
the central peak for $T<0.8$ K, indicating the onset of LRO. The spin-lattice
relaxation rate $1/T_{1}$ was measured in a temperature range $0.02$ K $\leq
T\leq300$ K. No clear indication of any kind of magnetic ordering was seen in
Sr$_{2}$Cu(PO$_{4}$)$_{2}$ and Ba$_{2}$Cu(PO$_{4}$)$_{2}$ from the $1/T_{1}$
data, whereas a clear indication of magnetic ordering was observed at
$\ T\approx0.8$ K ($k_{B}T/J\approx0.0074$) for BaCuP$_{2}$O$_{7}$. At
low-temperature, $1/T_{1}$ follows a nearly logarithmic increase for
$T\leq0.5$ K for Sr$_{2}$Cu(PO$_{4}$)$_{2}$ and Ba$_{2}$Cu(PO$_{4}$)$_{2}$,
which is expected for a 1D $S=1/2$ HAF system. Though the transverse decays
follow Gaussian behavior for all our samples, they result from dipole-dipole
interaction rather than a spinon mediated interaction. Our experimental
evidences on these compounds strongly reflect their low-dimensional nature
making them one of the best 1D $S=1/2$ HAF systems that have been looked at so far.

\begin{acknowledgments}
We would like to thank H.Alloul and J. Bobroff for some initial measurements
and helpful discussions. \ We thank H.$-$A.Krug von Nidda for EPR
measurements. \ One of us (AVM) would like to thank the Alexander von Humboldt
foundation for financial support for the stay at Augsburg. This work was
supported by the BMBF via VDI/EKM, FKZ 13N6917-A and by the Deutsche
Forschungsgemeinschaft (DFG) through the Sonderforschungsbereich SFB 484 (Augsburg).
\end{acknowledgments}

\textbf{Figure Captions}

FIG. 1 Schematic diagram of $\left[  \text{Cu(PO}_{\text{4}}\text{)}%
_{\text{2}}\right]  _{\infty}$ linear chains (a) propagating along
$b$-direction for (Ba/Sr)$_{2}$Cu(PO$_{4}$)$_{2}$ and (b) along the
$c$-direction for BaCuP$_{2}$O$_{7}$. The arrows show the direction of chains
while the possible interaction paths are shown by solid lines between atoms.

FIG. 2 X-ray diffraction pattern (Intensity vs. 2$\theta$) for (a) Sr$_{2}%
$Cu(PO$_{4}$)$_{2}$, (b) Ba$_{2}$Cu(PO$_{4}$)$_{2}$, and (c) BaCuP$_{2}$%
O$_{7}$. Miller indices are not shown in order to maintain the clarity of the
figures. Impurity peaks are marked by asterisks.

FIG. 3 Magnetic susceptibility ($M/H$) vs. temperature $T$ \ for (a) Sr$_{2}%
$Cu(PO$_{4}$)$_{2}$, (b) Ba$_{2}$Cu(PO$_{4}$)$_{2}$, and (c) BaCuP$_{2}$%
O$_{7}$ in an applied field of 5 kG. The solid lines are best fits of the data
to Eq. 1. Insets of (a) and (b) display the comparison of the magnetic
susceptibility for as-prepared and reduced samples\ at lower temperatures.
Susceptibility variations as a function of $\frac{1}{T}$ are shown to
demonstrate the lowering of Curie term on reduction.

FIG. 4 $^{31}$P shift $K$ vs. temperature $T$ for (a) Sr$_{2}$Cu(PO$_{4}%
$)$_{2}$, (b) Ba$_{2}$Cu(PO$_{4}$)$_{2}$, and (c) BaCuP$_{2}$O$_{7}$. The
solid lines are fits of Eq. 2 in the temperature range, $10$ K $\leq T\leq300$
K and then extrapolated down to $0.01$ K. Inset shows $K$ vs. $T$ on a
logarithmic \ temperature scale for improved visualisation of the low-$T$ data.

FIG. 5 $^{31}$P shift $K$ vs. spin susceptibility $\chi_{Spin}$ for
BaCuP$_{2}$O$_{7}$. The solid line shows the linear fit.

FIG. 6 Low-field ($H$ $\simeq$ $4$ kG) $^{31}$P NMR spectra at different
temperatures $T$ for (a) Sr$_{2}$Cu(PO$_{4}$)$_{2}$ and (b) Ba$_{2}$%
Cu(PO$_{4}$)$_{2}$. The insets of (a) and (b) contain the high-field and
low-field spectra at $1$ K. (c) and (d) show high-field and low-field spectra,
respectively, for BaCuP$_{2}$O$_{7}$ at various temperatures below $1$ K,
showing the sudden change in line width there.

FIG. 7 Spin-lattice relaxation rate $1/T_{1}$ (both high and low fields) vs.
temperature $T$ \ for (a) Sr$_{2}$Cu(PO$_{4}$)$_{2}$, (b) Ba$_{2}$Cu(PO$_{4}%
$)$_{2}$, and (c) BaCuP$_{2}$O$_{7}$. In the inset of (a), the high-field
magnetization recoveries are plotted as a function of pulse separation $t$ and
the solid line is a single-exponential fit for Sr$_{2}$Cu(PO$_{4}$)$_{2}$. The
inset of (b) displays the relaxation rate data for Ba$_{2}$Cu(PO$_{4}$)$_{2}$
on a linear temperature scale. In the inset of (c), the low-field
magnetization recoveries are plotted as a function of pulse separation $t$ at
$1$ K and $2$ K. The $1$ K data are fitted to double exponential (Eq. 3),
while $2$ K data are fitted to single exponential.

FIG. 8 Spin-echo decays are plotted as a function of $t^{2}$ at two different
temperatures for (a) Sr$_{2}$Cu(PO$_{4}$)$_{2}$, (b) Ba$_{2}$Cu(PO$_{4}$%
)$_{2}$, and (c) BaCuP$_{2}$O$_{7}$. The solid lines show the fitting to a
Gaussian function (Eq. 4). In the insets, $1/T_{2G}$ is plotted as a function
of temperature $T$.

FIG. 9 Distribution function $g(x)$ which represents the NMR spectrum is
plotted for (a) Cu site and (b) P sites. In the insets of (a) and (b),
$\left(  -1\right)  ^{l}\chi_{alt}\left(  l\right)  $ and $\left(  -1\right)
^{l}\chi_{alt}\left(  l\right)  +\left(  -1\right)  ^{l+1}\chi_{alt}\left(
l+1\right)  $ respectively are plotted\ as a function of the site index $l$
from the chain end at two different temperatures.

FIG. 10 Normalized experimental data (open symbols) as well as theoretical
curve (solid line) of $(1/T_{1})_{norm}$ are plotted vs. $k_{B}T/J$ for
Sr$_{2}$Cu(PO$_{4}$)$_{2}$, Ba$_{2}$Cu(PO$_{4}$)$_{2}$, and BaCuP$_{2}$O$_{7}$
at low temperatures, $0.1$ K $\leq T\leq10$ K. The data and the theoretical
curves have been scaled to $1$ at $10$ K.

\bigskip

\textbf{Table Captions}

TABLE I. Values of the parameters ($\chi_{0}$, $C$, $\theta$, and $J/k_{B}$)
obtained by fitting the bulk susceptibility to Eq. 1 to for each of the three samples.

TABLE II. Values of the parameters ($K_{0}$, $A$, and $J/k_{B}$) obtained by
fitting the NMR shift data to Eq. 2 in the temperature range, $10$ K $\leq
T\leq300$ K for each of the three samples.

\end{document}